\documentclass[10pt, conference, letterpaper]{IEEEtran}
\IEEEoverridecommandlockouts

\usepackage{cite}
\usepackage{amsmath,amssymb,amsfonts}

\usepackage{graphicx}
\usepackage{textcomp}
\usepackage{xcolor}
\usepackage{dsfont}

\usepackage{url}
\usepackage{algpseudocode}
\usepackage{multirow}
\usepackage{tablefootnote}
\usepackage{threeparttable}
\usepackage{caption}
\usepackage{subcaption}

\def\BibTeX{{\rm B\kern-.05em{\sc i\kern-.025em b}\kern-.08em
    T\kern-.1667em\lower.7ex\hbox{E}\kern-.125emX}}
\begin{document}

\title{MalMoE: Mixture-of-Experts Enhanced Encrypted Malicious Traffic Detection Under Graph Drift}

\author{
\IEEEauthorblockN{
Yunpeng Tan\textsuperscript{1},
Qingyang Li\textsuperscript{1},
Mingxin Yang\textsuperscript{1},
Yannan Hu\textsuperscript{2},
Lei Zhang\textsuperscript{2}\textsuperscript{*},
Xinggong Zhang\textsuperscript{1}\textsuperscript{*}
}
\IEEEauthorblockA{\textsuperscript{1}Peking University \textsuperscript{2}Zhongguancun Laboratory}
\thanks{\textsuperscript{*}Xinggong Zhang and Lei Zhang are the corresponding author (zhangxg@pku.edu.cn).}
}

\maketitle

\begin{abstract}
Encryption has been commonly used in network traffic to secure transmission, but it also brings challenges for malicious traffic detection, due to the invisibility of the packet payload. Graph-based methods are emerging as promising solutions by leveraging multi-host interactions to promote detection accuracy. But most of them face a critical problem: Graph Drift, where the flow statistics or topological information of a graph change over time. 


To overcome these drawbacks, we propose a graph-assisted encrypted traffic detection system, MalMoE, which applies Mixture of Experts (MoE) to select the best expert model for drift-aware classification. Particularly, we design 1-hop-GNN-like expert models that handle different graph drifts by analyzing graphs with different features. Then, the redesigned gate model conducts expert selection according to the actual drift. MalMoE is trained with a stable two-stage training strategy with data augmentation, which effectively guides the gate on how to perform routing. Experiments on open-source, synthetic, and real-world datasets show that MalMoE can perform precise and real-time detection.

\end{abstract}

\begin{IEEEkeywords}
encrypted traffic detection, cybersecurity, data drift, mixture of experts
\end{IEEEkeywords}

\section{Introduction}

Nowadays, network traffic encryption has been widely adopted to secure network transportation, but it also facilitates cyberattacks, which is an important issue in Network Observation. As reported in~\cite{SSL_Certificate_Authorities}, 97.4\% of the websites use the SSL certificate authorities, and the report of Zscaler~\cite{ThreatLabz_2024_Encrypted_Attacks_Report} shows that 87.2\% of malicious traffic is encrypted. These encrypted attacks render Deep Packet Inspection (DPI)~\cite{el2017survey,hu2007malware} ineffective, and their stealthy behaviors (e.g., C2 over DoH~\cite{bumanglag2020impact,patsakis2020encrypted}) cause methods based on single-flow statistics~\cite{ssl2015data,qin2015ddos,petliak2023signature} to fail as well. To address the issue above, some works~\cite{hsieh2024netvigil,lo2022graphsage,fu2023detecting,nguyen2023ts,duan2024practical,altaf2023ne,duan2022application} use graphs to analyze the interaction of flows to enhance detection accuracy. In flow graphs, nodes typically represent IP addresses, edges correspond to flows, node features capture the network contexts of IPs, and edge features reflect the statistical characteristics of the flows.


However, \textbf{\textit{temporal graph drift}} is a serious problem for graph-assisted encrypted traffic detection. It mainly stems from the variation of graph’s traffic statistics and topological information over time. Since most existing works are trained on traffic from a specific time window and evaluated on the subsequent traffic, graph drift can cause a severe distribution mismatch between the training and testing data. As shown in Section \ref{subsec:challenges}, we find two common types of temporal graph drift: \textbf{\textit{flow statistic drift}} (e.g., the number of bytes, the number of packets, and flow duration) caused by congestion; \textbf{\textit{graph scale drift}} caused by the variation of the number of connections. How to tackle the temporal graph drifts remains a challenge.


Most graph-based encrypted malicious detection methods~\cite{lo2022graphsage,nguyen2023ts,altaf2023ne,duan2022application} simply apply GNN for flow-level classification, ignoring the problem of temporal graph drift.
Anomaly detection works~\cite{luo2024identifying,fu2023detecting,hsieh2024netvigil,fu2021realtime} employ unsupervised learning to model benign traffic, but suffer from benign traffic drift, leading to high false positive rates. The works of model retraining~\cite{chen2025cd,xavier2024fast,hsieh2024netvigil} need to periodically finetune the detection model, but it is impractical in the real world due to the large consumption of training resources (e.g., memory, time, and training labels).  
The works with data augmentation~\cite{deng2024robust,hsieh2024netvigil,xie2023rosetta} apply data augmentation to train the model, but a single model is incapable of handling all kinds of graph drifts due to its limited representation capacity.

Thus, we aim to find a \textbf{retraining-free} method that could \textbf{combine the benefits of different model designs} for \textbf{real-time} encrypted malicious traffic detection. Luckily, through experiments detailed in Section \ref{subsec:motivation}, we find that \textbf{different node feature types are inherently robust to different graph drifts}. As an example, in this paper, we consider two types of node features: \textbf{\textit{average traffic feature}} (robust to graph scale drift) and \textbf{\textit{node degree feature}} (robust to flow statistic drift). Through supervised training, they can inherently process their corresponding drifts without any retraining. However, simply using the concatenation of these node features for traffic detection does not work, because both of them are sensitive to the other type of graph drift. For example, the average traffic feature fails under flow statistic drift, and the node degree feature fails under graph scale drift.

Inspired by this insight and a classic technique that can combine the power of different models: Mixture-of-Experts (MoE)~\cite{jacobs1991adaptive}, we propose \textbf{\textit{MalMoE}}, a graph-drift-aware encrypted malicious traffic detection system that uses MoE to combine different drift resistances of different node feature types. In MalMoE, each expert handles a specific type of node feature, thereby exhibiting robustness to a specific type of graph drift. The gate takes the original inputs of the experts and outputs the results of expert selection, thereby integrating the predictions of experts to produce the final prediction. MalMoE is a highly extensible method, as new experts can be flexibly added to handle new types of graph drift.

To validate our insight, we encounter three technical challenges: \textbf{First}, GNN is known to hold low efficiency~\cite{fu2023detecting} and usually handles node-level information, but in traffic detection, we have to handle edge-level classification with high efficiency. \textbf{Second}, the gate design of the traditional MoEs does not suit our scenario. On the one hand, using only sample-level information like traditional MoEs is not enough to distinguish near-OOD (Out-of-Distribution) samples (i.e., data points that are slightly different from the training distribution). On the other hand, the traditional "weighted summation" mechanism increases the training difficulty and harms the explainability of the gating weights. \textbf{Third}, existing datasets are not suitable for training MoE. They usually show few temporal graph drifts, rendering the gating selection redundant. Moreover, as training progresses, fluctuations in expert performance can lead to instability in the training of the gate.

Thus, we propose three solutions for these challenges: \textbf{For the first challenge}, we design a simple but effective 1-hop-GNN-like expert network, where each expert utilizes a specific node feature type for edge classification. \textbf{For the second challenge}, we redesign the gate model from two aspects to improve its performance on expert routing. On the one hand, we additionally input the graph representations into the gate model to detect near-OOD samples. On the other hand, we train the gate model with a hard selection task to improve its generalizability and explainability. \textbf{For the third challenge}, we use data augmentation to guide the gate in making decisions under different drifts, and apply two-stage training to ensure stable gate training.

We evaluate MalMoE on public datasets, a synthetic dataset, and real-world traces, where MalMoE can outperform the baselines by at least 24\% in ACC and 31\% in F1 under graph drifts. Besides, after simple optimization, MalMoE can perform detection at 858,646 flows/s, which satisfies the requirements of real-time detection.

We summarize our contributions as follows:

\begin{itemize}
    \item We propose MalMoE, a graph-based encrypted malicious traffic detection system focusing on the problem of temporal graph drifts. As we know, we're the first to use MoE to integrate drift resistances of different node features for robust traffic detection.
    \item We design 1-hop-GNN-like expert models to effectively conduct edge classification under graph drifts.
    \item We adjust the gate design to improve its routing performance in drift-aware detection.
    \item We apply a two-stage training strategy with data augmentation, to enable effective and stable training.
    \item We apply datasets from various sources to evaluate the effectiveness and efficiency of MalMoE.
\end{itemize}

\section{Background and Motivation}

\subsection{Challenges of Graph-Aided Encrypted Traffic Detection}
\label{subsec:challenges}


\begin{figure}
    \centering
    \includegraphics[width=0.9\linewidth]{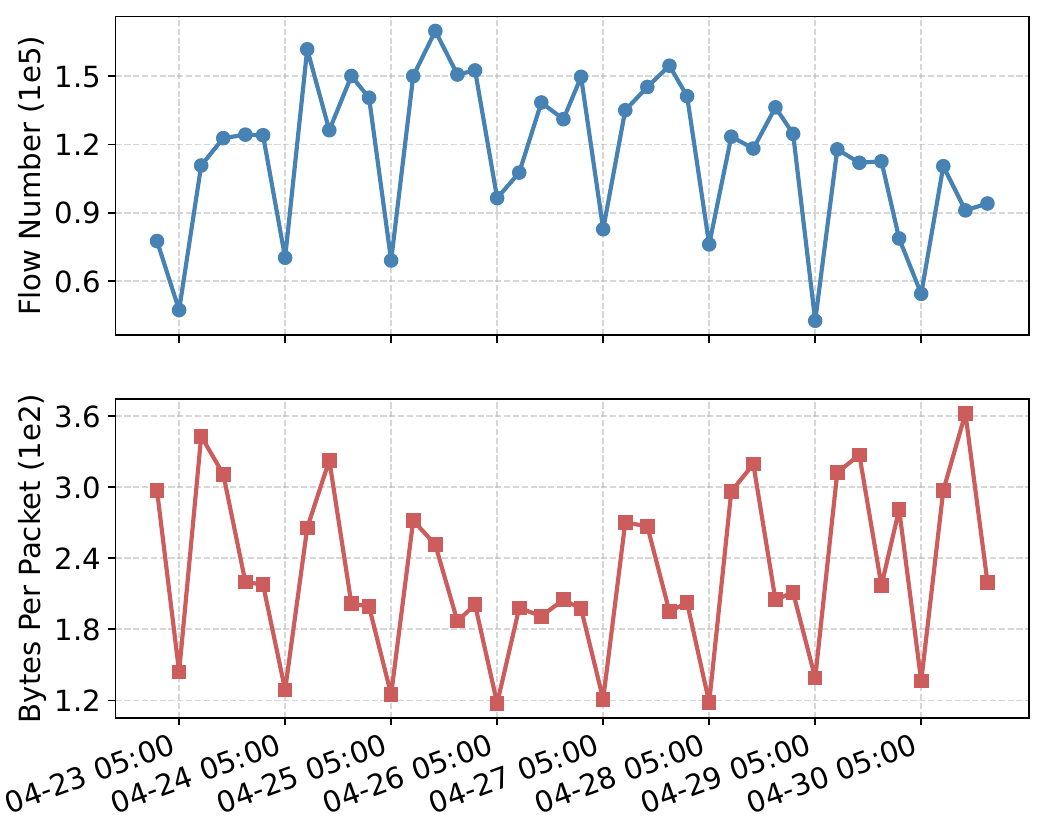}
    \caption{Drifts observed in the "flow number" and "bytes per packet" of a well-known backbone network operator.}
    \label{fig:challenge}
\end{figure}

The frequent encrypted cyber attacks usually exhibit stealthy behaviors as benign traffic, which has given rise to many graph-based approaches~\cite{hsieh2024netvigil,lo2022graphsage,fu2023detecting,nguyen2023ts,duan2024practical,altaf2023ne,duan2022application}. However, the graph-based methods always face a severe problem: temporal graph drift, which means that the topological or statistical features of the graph vary along with time. Through analysis of real-world traces from a well-known backbone network operator, we find two kinds of common drifts, namely \textbf{\textit{flow statistic drift}} and \textbf{\textit{graph scale drift}}, as shown in Figure \ref{fig:challenge}. We analyze the traffic from 2025-04-23 00:00 to 2025-04-30 20:00, and collect the instantaneous (within 30 seconds) flow numbers (representing graph scale) and bytes per packet (representing flow statistic). It can be seen that both quantities exhibit periodicity, reaching the minimum value around 05:00 (resting time) and the maximum value around 15:00 (working time), and the ratio between the maximum and minimum values can reach 400\% for flow number and 300\% for bytes per packet. It illustrates obvious temporal graph drifts. Besides, we also prove later in Section~\ref{subsec:motivation} that the two kinds of drifts can severely deteriorate the performance of graph-based analysis.

There have been lots of works focusing on the drifting problem, but they all have certain drawbacks. Some works~\cite{luo2024identifying,fu2023detecting,hsieh2024netvigil,fu2021realtime} use unsupervised learning to detect unknown attacks, which learn the pattern of benign flows and treat the abnormal flows as malicious flows. But these methods fail when the distribution of benign traffic drifts as well, which is common because user behaviors are easy to vary along with time. Some works~\cite{chen2025cd,xavier2024fast,hsieh2024netvigil} employ periodical retraining, but the labels of the retraining data are usually expensive to get, since it requires network security experts for labeling. The other works~\cite{deng2024robust,hsieh2024netvigil} use data augmentation during training, enforcing a single model to handle all graph drifts. It is impractical because a single model only has limited expressiveness, and it's impossible for it to remember all kinds of data distributions.

\subsection{Threat Model and Design Goals}
\label{subsec:threat}

\begin{figure}
    \centering
    \includegraphics[width=0.7\linewidth]{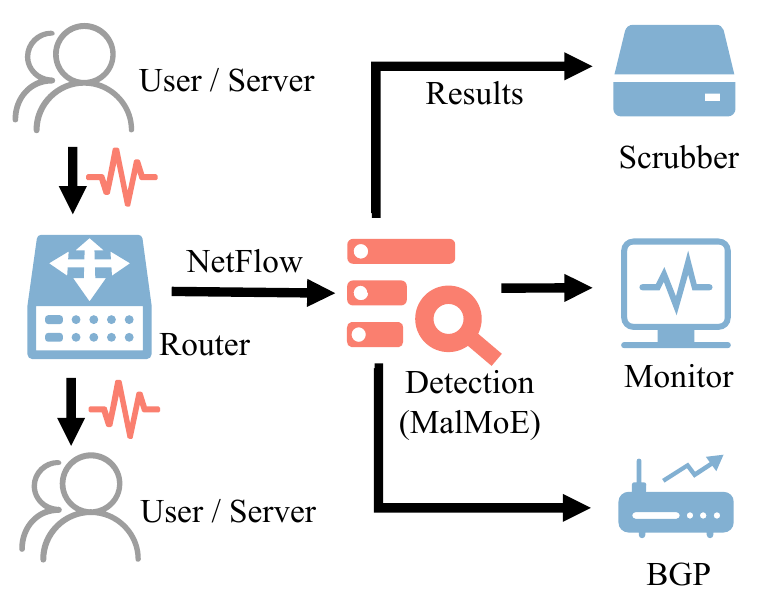}
    \caption{Threat model of MalMoE's flow-level detection.}
    \label{fig:system}
\end{figure}

Our objective is to detect encrypted malicious traffic under temporal graph drift, using flow-level information from the router as input. As shown in Figure \ref{fig:system}, when the network traffic passes through the router, it aggregates the packet information into flow statistics in the format of NetFlow~\cite{netflow_documentation} or NetStream~\cite{netstream_web}, and reports the NetFlow data to the detection system (i.e., MalMoE). On the detection device, MalMoE analyzes both the topological and statistical information (which may vary along with time), and figures out the malicious flows. The detection results can be sent to various downstream applications, such as the scrubbing device, the traffic monitoring device, and the BGP for route blocking.

There are three design goals for MalMoE:

\begin{itemize}
    \item \textbf{Retraining-free.} It should depend on the inherent robustness of the model design to avoid the costly labeling process required by retraining.
    \item \textbf{Extensible.} Its framework should support combining the robustness of different detection models, to avoid being limited by the detection power of a specific model.
    \item \textbf{Real-time.} It should perform real-time detection on consumer-grade devices, which means that the malicious flows should be detected as soon as the NetFlow data is reported from the router.
\end{itemize}

\subsection{Insight for using MoE}
\label{subsec:motivation}
\begin{figure}
\includegraphics[width=\linewidth]{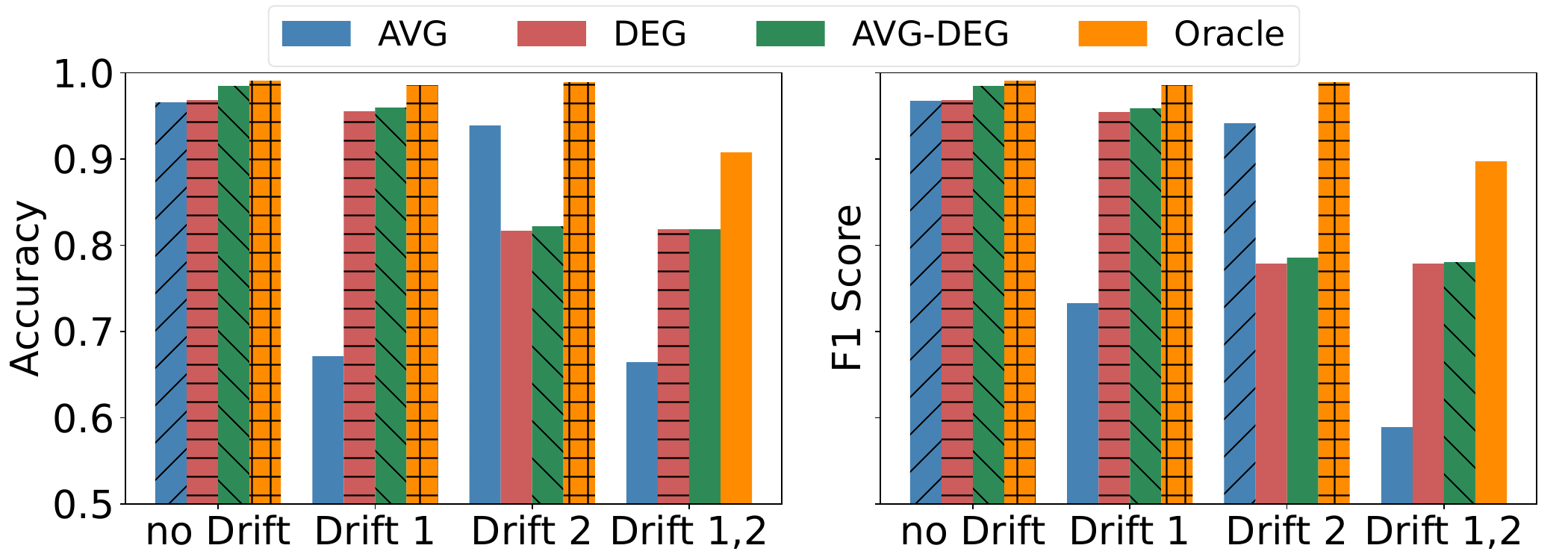}
\caption{Accuracy and F1 score measured on test graphs with different graph drift, using different node features.}
\label{fig:insight}
\end{figure}

To settle the graph drifts, we find that the way to extract node features can inherently handle the drifts without any retraining. For the two graph drifts: flow statistic drift (Drift 1) and graph scale drift (Drift 2), we find that two kinds of node features: average traffic feature (AVG) and node degree feature (DEG) can handle them separately. We apply these node features to train a GNN, and show the results in Figure \ref{fig:insight}. In detail, "AVG" uses the average of the statistics of neighboring flows of a node to represent the node, which is inherently robust to graph scale drifts; "DEG" uses the number of neighboring flows of a node to represent the node, which is inherently robust to flow statistic drifts. However, they always fail under the other types of drift.

Thus, we are wondering, \textbf{is it possible to combine their robustness to different graph drifts?} Firstly, we try an intuitive way by using the concatenation of the two kinds of node features as the input (AVG-DEG), but it shows two problems: First, it may bias towards the node feature with larger numerical range. In Figure \ref{fig:insight}, it can be seen that "AVG-DEG" shows similar results to "DEG"; Second, under different drifts, some part of the concatenated features always drift, which may even deteriorate the performance of the model. Thus, we try another way: for each flow, we choose to trust the prediction from the model trained with "better" node features. We design an oracle algorithm, which always chooses the model with lower classification loss (i.e., per-sample Cross Entropy). As shown in Figure \ref{fig:insight}, under different circumstances, it can greatly surpass the other methods. However, we do not know the ground truth labels during inference. \textbf{How to automatically select the node feature for each flow to be detected} becomes the focus of the remaining paper.

\section{Method}
\label{sec:method}

\subsection{Overview}
\label{subsec:overview}

We present MalMoE, a retraining-free, extensible, and real-time flow-level traffic detection method with the help of graph analysis, particularly designed for encrypted traffic under temporal graph drift. Our insight is that different node features inherently remain stable under different graph drifts. Inspired by this observation, we combine their resistances to different graph drifts using a Mixture of Experts (MoE) architecture, where each expert applies one node feature type to handle one kind of graph drift. Since the node features are inherently robust, the experts are tolerant to graph drifts without retraining. The MoE architecture also enables flexible combination of different types of node features and detection models. Due to the simplicity of our model design, it can detect malicious flows as soon as the router reports.

\begin{figure}
    \centering
    \includegraphics[width=0.9\linewidth]{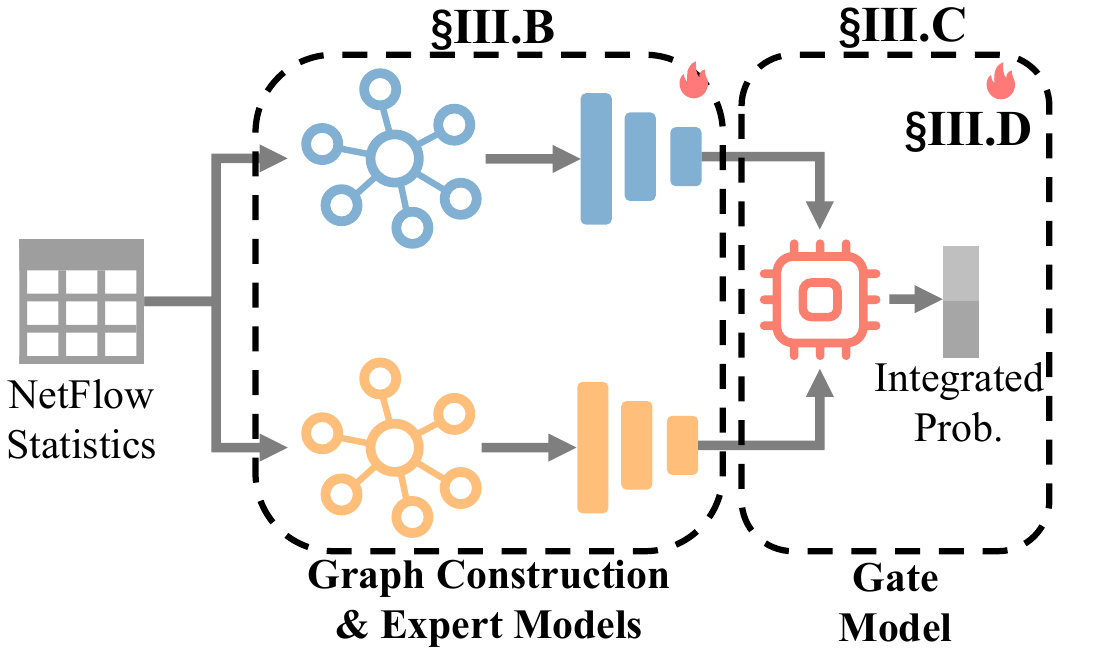}
    \caption{The overview of MalMoE.}
    \label{fig:overview}
\end{figure}

Figure~\ref{fig:overview} gives an overview of MalMoE, which mainly consists of three parts: The flow statistics are first constructed into a flow graph where the nodes have different types of features, which will be analyzed individually by our simple but effective GNN-like expert models (Section \ref{subsec:construction_expert}), to ensure the efficiency of edge-level classification. Then, a gate model automatically integrates the results from experts to form the final prediction. To promote the performance of the gate model, we make two improvements to it: adding graph representation as input and applying hard selection (Section \ref{subsec:gate}). To put them together, we apply a two-stage supervised training strategy with data augmentation, to ensure the effective and stable training of the model (Section \ref{subsec:train}).

\subsection{Graph Construction and Expert Models}
\label{subsec:construction_expert}

In this section, as shown in Figure \ref{fig:expert}, we will discuss how we transform the collected flow-level statistics into a traffic graph (Section \ref{subsubsec:construction}) to enable graph analysis, and how the expert models process the graph in a simple but effective way to conduct edge-level classification(\ref{subsubsec:expert}).

\subsubsection{Graph Construction}
\label{subsubsec:construction}

The input of MalMoE is pieces of flow statistics (such as NetFlow). To analyze flow interactions, they are transformed into a graph with edge features and node features. In this graph, each node stands for an IP, and each edge stands for a network flow. Considering a piece of NetFlow record, it is represented by an edge $(u,v)$ between its source IP $\mathrm{IP}_{u}$ and its destination IP $\mathrm{IP}_{v}$, with the flow statistics as the edge feature $f_{(u,v)}$, (e.g., in/out bytes and in/out packets). It's noticeable that there may be multiple flows from $\mathrm{IP}_u$ to $\mathrm{IP}_{v}$, so there may be multiple edges between node $u$ and node $v$. \textbf{For simplicity, we still use $(u,v)$ to represent an edge (i.e., a single flow) instead of an IP pair (which may contain several flows) in the following paper.}

There are different ways to extract node features. Here, we use two types of features as representatives: \textbf{\textit{average traffic feature (avg)}} and \textbf{\textit{node degree feature (deg)}}. For a node $u$, if the nodes connecting to $u$ are $N_{u}^{in}$, and the nodes that $u$ connects to are $N_{u}^{out}$, the average traffic feature $h_{u,avg}$ and node degree feature $h_{u,deg}$ are calculated as:
\begin{equation}
h_{u,avg} = \frac{\sum_{i \in N_{u}^{in}}f_{(i,u)} + \sum_{j \in N_u^{out}}f_{(u,j)}}{|N_{u}^{in}|+|N_{u}^{out}|},
\end{equation}
\begin{equation}
h_{u,deg}=|N_{u}^{in}|+|N_{u}^{out}|.
\end{equation}
It can be seen that $h_{u,avg}$ is gained by averaging the edge features, and $h_{u,deg}$ is the degree of node $u$.

We have generated a graph $G=\{N, E, H_{avg},H_{deg},F\}$, where $N=\{u\}$ is the node set, $E=\{(u,v)\}$ is the edge set, $H_{avg}=\{h_{u,avg}\}$ and $H_{deg}=\{h_{u, deg}\}$ are average traffic node features and degree node features, and $F=\{f_{(u,v)}\}$ are edge features. Our task is, given a set of training graphs $\{G_{train}\}$ and their edge labels $\{\mathrm{Y}_{train}\}$ (whether a flow is benign/malicious), we need to predict the edge labels of testing graphs $\{G_{test}\}$, and there may exist graph drifts between $\{G_{train}\}$ and $\{G_{test}\}$.

\subsubsection{Expert Models}
\label{subsubsec:expert}

\begin{figure}
    \centering
    \includegraphics[width=0.9\linewidth]{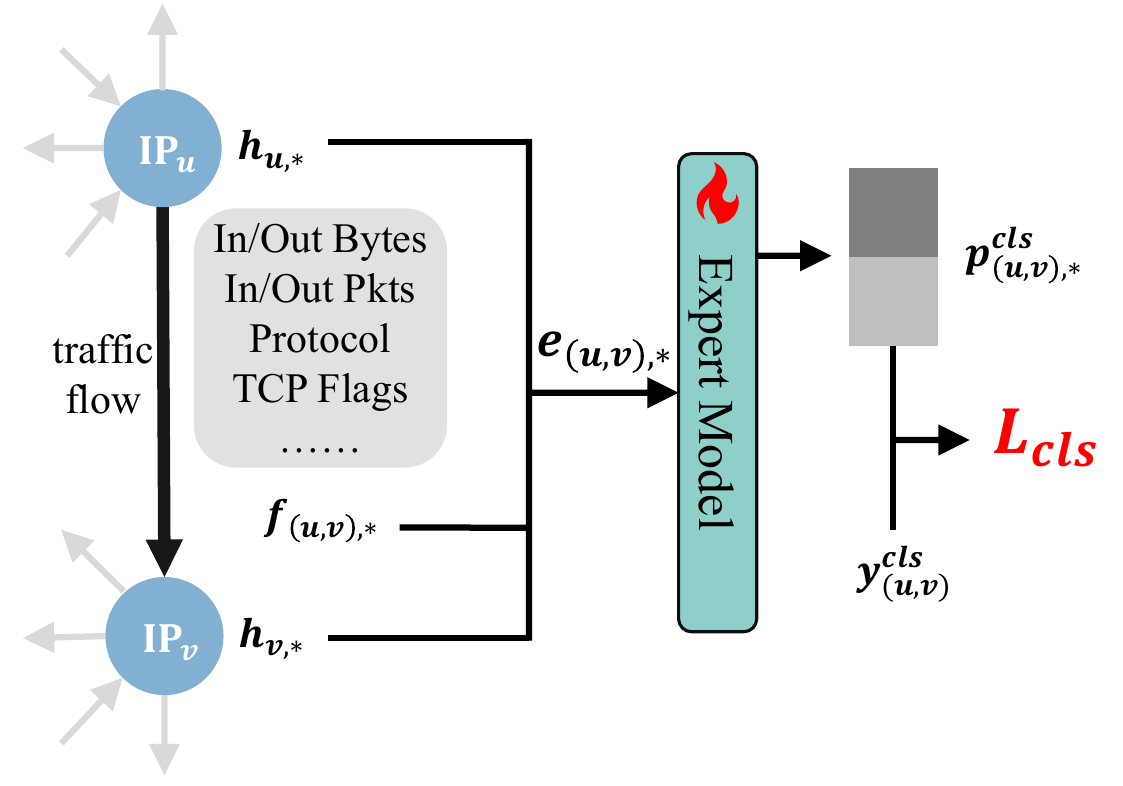}
    \caption{The input and architecture of the expert models.}
    \label{fig:expert}
\end{figure}

Existing GNNs mostly conduct graph-level/node-level classification and incur high computational cost~\cite{fu2023detecting}, but our scenario requires efficient edge-level classification. According to the discussion in NetVigil~\cite{hsieh2024netvigil}, a one-hop GNN is sufficient to capture the essential structural information in the network traffic graph, because in most cases, the router can only capture the traffic directly from the attackers to the protected targets. Thus, for simplicity, we use a similar structure to one-hop GNN: We view the extracted node features as one-hop information, and directly use node features with edge features for classification. 

With the constructed graphs, we build two expert models - \textbf{\textit{"Avg-Expert"}} (handling average traffic feature) and \textbf{\textit{"Deg-Expert"}} (handling node degree feature) - to process different kinds of node features separately. Specifically, as illustrated in Figure \ref{fig:expert}, we use "*" to represent a certain kind of node features ("avg" or "deg"). For a flow from $\mathrm{IP}_u$ to $\mathrm{IP}_v$, we concatenate the node features and the edge feature as the flow embedding $e_{(u,v),*}=h_{u,*} || h_{v,*} ||f_{(u,v), *}$. The flow embedding is input into the corresponding expert model, which is actually a simple MLP with parameters $\theta_{cls}^{*}$, to get the predicted probability of the flow:
\begin{equation}
p_{(u,v),*}^{cls}=\mathrm{MLP}(e_{(u,v),*};\theta_{cls}^{*}).
\end{equation}
Then, we calculate the per-flow classification loss (Cross Entropy) $L_{(u,v), *}^{cls}$, which will be used in the gate model:
\begin{equation}
L_{(u,v),*}^{cls}=\mathrm{CE}(p_{(u,v),*}^{cls},y_{(u,v)}).
\end{equation}
The classification loss used for training is the average of the per-flow losses:
\begin{equation}
L_{cls}=\frac{1}{|E|}\sum_{(u,v)\in E}(L_{(u,v),avg}^{cls}+L_{(u,v),deg}^{cls}).
\end{equation}

Due to the nature of different node features, different experts show robustness to different graph drifts. As for the Avg-Expert, it shows robustness to node degree drift. For example, the traffic at midnight is usually smaller than that during the day. However, as long as the distribution of flow statistics doesn't vary a lot, the average traffic feature shows robustness. As for the Deg-Expert, it is robust to flow statistic drift. For example, changes in network bandwidth can affect statistics such as the throughput of bytes, the throughput of packets, and the average inter-packet interval, but the number of connections might remain stable.

\subsection{Gate Model}
\label{subsec:gate}

\begin{figure}
    \centering
    \includegraphics[width=0.9\linewidth]{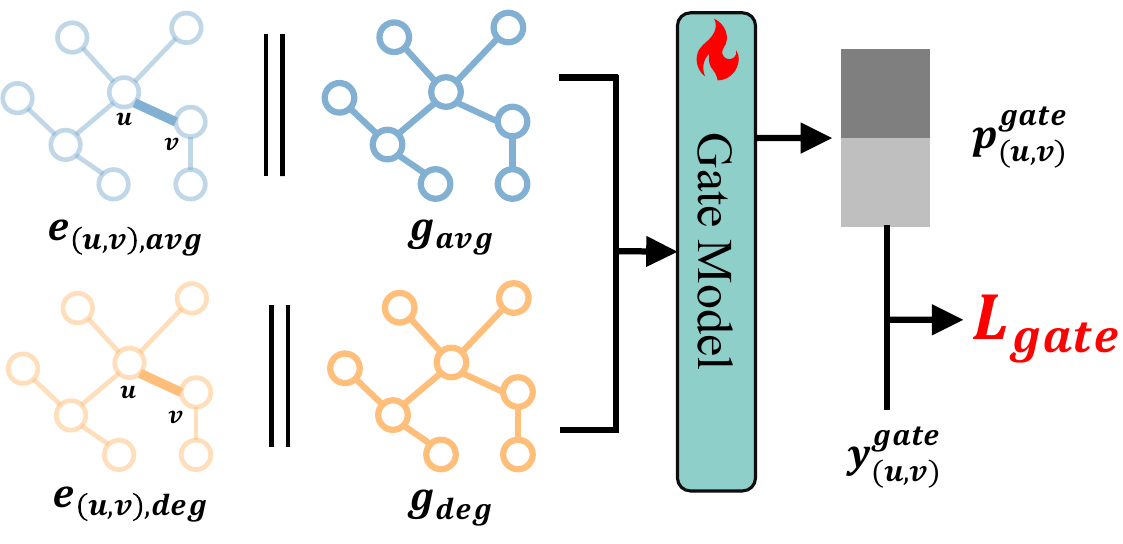}
    \caption{The input and architecture of the gate model.}
    \label{fig:gate}
\end{figure}

Since the expert models show different abilities, the gate model aims to integrate them by selecting experts for each flow, as shown in Figure \ref{fig:gate}. It is attainable because for each flow, there is always one type of node feature with the least degree of drift, which can prompt the gate to trust the corresponding expert. However, traditional gating designs are not well suited to our scenario, due to their limited receptive field (Section \ref{subsubsec:gate_input}) and the "weighted summation" gating mechanism (Section \ref{subsubsec:gate_output}), so we will discuss our improvements below.

\subsubsection{Gate Model Input}
\label{subsubsec:gate_input}


Traditional MoEs use only per-sample features for routing, which is insufficient for drift-aware traffic detection. In classic MoEs, experts typically correspond to fixed domains~\cite{jiang2024med}, so a single sample often contains enough cues to determine its domain. However, per-sample information alone is not enough to determine whether the sample has drifted. For example, a Deg-Expert trained on daytime (high-volume) traffic may miss early-morning (low-volume) attacks because malicious-node degrees can fall into the daytime benign range, causing false negatives. However, such near-OOD samples can mislead the gate to choose Deg-Expert instead of Avg-Expert, producing wrong predictions.

We find that integrating graph-level information can settle this problem. Following the above example, although the flow's node degree may appear to be near-OOD, the traffic graphs in the early morning are significantly smaller than the daytime, which can be a sign for the gate to avoid selecting the Deg-Expert. The process of extracting graph representation $g_{*}$ is usually called "Readout" in GNN:
\begin{equation}
g_{*} = \mathrm{Readout}(\{e_{(u,v),*}\}_{(u,v)\in E}).
\end{equation}
Here we apply the simplest readout: $g_{*}=\mathrm{Mean}(\{e_{(u,v),*}\})$. After that, the flow embedding and the graph embedding are concatenated as a whole as the input of the gate model:
\begin{equation}
p_{(u,v)}^{gate}=\mathrm{MLP}(g_{avg}||g_{deg};\theta_{gate}),
\end{equation}
where $\theta_{gate}$ is the parameters of our MLP-based gate model. The output of the gate model is the probability of assigning each sample to each expert.

\subsubsection{Gate Model Output}
\label{subsubsec:gate_output}

Usually, the gate model outputs weights to sum up the results from each expert (usually the latent vectors of the expert models), which brings two limitations under our scenario:

\begin{itemize}
    \item \textbf{It makes the training more challenging.} This manner needs to train a classification head to process the aggregated latent vector, which requires strong generalizability as the gate model. However, for traffic detection, the labels are hard to get, leading to a small training dataset, making such generalizability hard to achieve.
    \item \textbf{The assigned weights lack explainability.} The latent vectors of different experts lie in different latent spaces, and the spaces also have different scales, which makes the weights generated by the gate lose their practical meaning. For example, due to the obviously larger latent vectors of the Deg-Expert, the gate always assigns tiny weights for the Deg-Expert. However, we hope the weights can reflect the dominant drift during inference.
\end{itemize} 

Shifting the gate's role from weighting expert outputs to selecting an expert in a "hard" manner for each sample can resolve these issues. Specifically, for a flow $(u,v)$ to be classified, at first, we construct a "gating label" $y_{(u,v)}^{gate}$:
\begin{equation}
y_{(u,v)}^{gate}=\mathrm{argmin}(L_{(u,v),avg}^{cls},L_{(u,v),deg}^{cls}),
\end{equation}
which means that the gate should select the expert with the smallest classification loss. However, when the experts both output correct or wrong predictions, the gate's selection makes no difference, so we calculate a $\mathrm{mask}$ to exclude such samples:
\begin{equation}
\mathrm{mask}_{(u,v)}=\mathbb{1}(\hat{y}_{(u,v),cls}\neq \hat{y}_{(u,v),avg}).
\end{equation}
When the experts make different predictions, the mask is set to 1. The training loss for the gate $L_{gate}$ is the sum of the classification loss (Cross Entropy) calculated with "gating labels", only considering masked flows:
\begin{equation}
L_{gate}=\frac{1}{||\mathrm{mask}||_1}\sum_{(u,v)\in E}\mathrm{CE}(p_{(u,v)}^{gate},y_{(u,v)}^{gate})\mathrm{mask}_{(u,v)}.
\end{equation}

\subsection{Training of MalMoE}
\label{subsec:train}

In MalMoE, there are two parts that need training: the expert models and the gate model. To enable the gate to achieve robustness from a relatively small training set, we use drift-motivated data augmentation (Section \ref{subsubsec:data_augmentation}). Besides, to avoid the problem of unstable training of the gate, we use a two-stage training strategy (\ref{subsubsec:two_stage_training}).

\subsubsection{Data Augmentation}
\label{subsubsec:data_augmentation}
Most existing datasets for flow-level traffic detection hold tiny temporal drift, but as discussed above, the gate model needs to achieve strong generalizability. Thus, we apply data augmentation on the training graphs to teach the gate how to select experts under different graph drifts. Specifically, we apply two kinds of data augmentation: flow statistic perturbation and random edge dropping, with respect to the two types of graph drifts we focus on. Given a graph $G(N,E,F)$, we can apply flow statistic perturbation:
\begin{equation}
F'=F+\epsilon+b,\ \epsilon\sim \mathcal{N}(0, \sigma),
\end{equation}
\begin{equation}
b\sim\mathcal{U}(-\alpha,\alpha),\ \sigma\sim\mathcal{U}(0,\beta),
\end{equation}
where $\alpha$ and $\beta$ are hyperparameters, $b$ and $\sigma$ are sampled from uniform distributions, and $\epsilon$ is a random vector following a normal distribution $\mathcal{N}(0, \sigma)$. We intentionally use a biased distribution to increase the degree of drift in the augmented graph. Also, we can apply random edge dropping:
\begin{equation}
E'=\{(u,v)\in E|m_{(u,v)}\le a\},\ m_{(u,v)}\sim \mathcal{U}(0,1),
\end{equation}
\begin{equation}
a\sim\mathcal{U}(\gamma,1),
\end{equation}
where $\gamma$ is a hyperparameter, $a$ is sampled from a uniform distribution, and $m_{(u,v)}$ is a random variable following a uniform distribution. It means that we drop edges with a probability of $1-a$. In implementation, the two kinds of augmentation are used together to get the augmented graph $G'=\{N, E', F'\}$.

\subsubsection{Two-Stage Training}
\label{subsubsec:two_stage_training}

Traditional MoEs are usually trained in an end-to-end manner, but the performance of experts varies along with training, which may result in unstable gate training. Thus, we train MalMoE in two stages. At the first stage, we only use $L_{cls}$ to train the expert models, using one set of hyperparameters $(\alpha_1, \beta_1, \gamma_1)$. At the second stage, we only use $L_{gate}$ to train the gate model, using another set of hyperparameters $\{\alpha_{2},\beta_{2},\theta_2\}$, which satisfie: $\alpha_{2} \ge \alpha_{1}, \beta_{2}\ge\beta{1}, \gamma_{2}\ge\gamma_{1}$. This guarantees that the second stage applies stronger augmentation. Because if the experts both achieve excellent performance during the first stage (in that case, it doesn’t matter which expert the gate selects), we must ensure their performance differs in the second stage, so that the gate model can be trained.

Note that although we use data augmentation, we do not rely on a single model for detection like~\cite{deng2024robust,hsieh2024netvigil,xie2023rosetta}. For expert models, the robustness essentially stems from node features rather than data augmentation, as shown in Section \ref{subsec:effectiveness}. For the gate model, though it has to handle various drifts, its task only focuses on assigning experts, which is relatively easy compared with direct benign/malicious classification.

\section{Experimental Evaluation}

\setlength{\tabcolsep}{1.5pt}  
\begin{table*}[!t]
\caption{Results of different methods on different datasets, with and without graph drifts.}
\label{tab:real}
\centering
\begin{threeparttable}
\resizebox{\textwidth}{!}
    {
    \renewcommand{\arraystretch}{1.2}
    \begin{tabular}{c|c|c|c|c|c|c|c|c|c|c|c|c|c}
        \hline
         & & \multicolumn{2}{c}{CIC-IDS2018} & \multicolumn{2}{|c}{ToN-IoT} & \multicolumn{2}{|c}{BoT-IoT} & \multicolumn{2}{|c}{UNSW-NB15} & \multicolumn{2}{|c}{Synthetic} & \multicolumn{2}{|c}{Overall}\\
         \hline
        Method & Metric & \textcolor{blue}{w/o Drift} & \textcolor{red}{w/ Drift} & \textcolor{blue}{w/o Drift} & \textcolor{red}{w/ Drift} & \textcolor{blue}{w/o Drift} & \textcolor{red}{w/ Drift} & \textcolor{blue}{w/o Drift} & \textcolor{red}{w/ Drift} & \textcolor{blue}{w/o Drift} & \textcolor{red}{w/ Drift} & \textcolor{blue}{w/o Drift} & \textcolor{red}{w/ Drift}\\
         \hline
         \multirow{2}{*}{NetBeacon} & ACC & 0.9847 & 0.5458 & 0.8914 & 0.4335 & 0.9844 & 0.7249 & 0.9840 & 0.3776 & 0.7290 & 0.5632 & 0.9147 & 0.5290\\
         & F1 & 0.9815 & 0.2003 & 0.8895 & 0.3873 & 0.7526 & 0.0943 & 0.9919 & 0.5446 & 0.7540 & 0.3280 & 0.8739 & 0.3109\\
         \hline
         \multirow{2}{*}{E-GraphSAGE} & ACC & 0.9684 & 0.7937 & 0.8942 & 0.4423 & 0.9888 & 0.9783 & 0.9866 & 0.6805 & 0.9704 & 0.6911 & 0.9617 & 0.7172\\
         & F1 & 0.8796 & 0.6397 & 0.9057 & 0.4655 & 0.8691 & 0.6248 & 0.9932 & 0.8087 & 0.9710 & 0.6205 & 0.9237 & 0.6318\\
         \hline
         \multirow{2}{*}{HyperVision} & ACC & 0.7602 & 0.6029 & 0.7206 & 0.7542 & 0.9514 & 0.6591 & 0.8663 & 0.9945 & 0.6354 & 0.5478 & 0.7868 & 0.7117\\
         & F1 & 0.6861 & 0.5751 & 0.7134 & 0.8469 & 0.5345 & 0.1119 & 0.9275 & 0.9973 & 0.6391 & 0.6045 & 0.7001 & 0.6271\\
         \hline
         \multirow{2}{*}{NetVigil} & ACC & 0.9029 & 0.6240 & \textbf{0.9437} & 0.6831 & 0.9528 & 0.4258 & 0.6392 & 0.0064 & 0.7189 & 0.5567 & 0.8315 & 0.4592\\
         & F1 & 0.9085 & 0.6107 & \textbf{0.9518} & 0.7669 & 0.5723 & 0.0790 & 0.7757 & 0.0000 & 0.7604 & 0.5990 & 0.7937 & 0.4111\\
         \hline
         \multirow{2}{*}{MalMoE} & ACC & \textbf{0.9970} & \textbf{0.9993} & 0.8784 & \textbf{0.9011} & \textbf{0.9977} & \textbf{0.9947} & \textbf{0.9967} & \textbf{0.9964} & \textbf{0.9880} & \textbf{0.9376} & \textbf{0.9716} & \textbf{0.9658}\\
         & F1 & \textbf{0.9933} & \textbf{0.9978} & 0.8987 & \textbf{0.9181} & \textbf{0.9608} & \textbf{0.8957} & \textbf{0.9983} & \textbf{0.9982} & \textbf{0.9881} & \textbf{0.9298} & \textbf{0.9678} & \textbf{0.9479}\\
         \hline
    \end{tabular}}
\end{threeparttable}
\end{table*}

\subsection{Experiment Setup}
\label{subsec:implementation_details}

\subsubsection{Dataset}

Since the proposed MalMoE can be applied to both encrypted and unencrypted network traffic as long as only the flow-level information is used, we can conduct experiments using flow-level datasets.

To evaluate the performance of MalMoE, we conduct experiments on 4 public datasets, namely NF-CSE-CIC-IDS2018-v3 (12.93\% malicious), NF-ToN-IoT-v3 (38.98\% malicious), NF-UNSW-NB15-v3 (5.4\% malicious), and NF-BoT-IoT (99.7\% malicious) from ~\cite{luay2025NetFlowDatasetsV3}. For each attack within each dataset, we split the flows into training and testing flows according to the arrival time of the flows. We also manually apply graph drifts to the testing graphs to simulate larger drift magnitudes.

Besides, to better learn the robustness of MalMoE to different graph drifts, we construct a synthesized dataset (50\% malicious) from the Infiltration attack of NF-CSE-CIC-IDS2018-v1~\cite{sarhan2021netflow} where individual flows lack discriminative features. We split flows into train/test, pre-augment training graphs with one parameter set for model training, and pre-augment test graphs with a different set to ensure a different distribution.

To illustrate the practical use of MalMoE, we test on real-world traces from a well-known backbone network operator. We extract NetStream samples from different dates and different clock times to show the robustness of MalMoE along with time. Before experiments, we use IXP Scrubber~\cite{wichtlhuber2022ixp} to label the flows, resulting in an average of 1\% of the traffic being identified as malicious.

To evaluate the efficiency of MalMoE, we use the MAWI~\cite{mawi_documentation} dataset, which contains 6,268,116 flows in total. We input the entire dataset into the model at once to simulate a high-throughput detection scenario.

\subsubsection{Baselines}

For the evaluation under different attacks, we use the following baselines:
\begin{itemize}
  \item \textbf{NetBeacon}~\cite{zhou2023efficient}: It applies Random Forest for supervised detection, using solely single-flow statistics.
  \item \textbf{E-GraphSAGE}~\cite{lo2022graphsage}: It trains a Graph Convolution Network in a supervised manner to detect malicious flows.
  \item \textbf{HyperVision}~\cite{fu2023detecting}: It designs an outlier detector for encrypted traffic detection with analysis of the traffic graph. We finetune the weights in its loss function to improve its performance.
  \item \textbf{NetVigil}~\cite{hsieh2024netvigil}: It is a GNN-based unsupervised malicious flow detection method designed for east-west traffic. To ensure fairness, we modify it to fit flow-level analysis like MalMoE, instead of IP-pair-level analysis.
\end{itemize}
Both HyperVision and NetVigil require a threshold to distinguish abnormal flows. We use the threshold that maximizes $\mathrm{TPR}-\mathrm{FPR}$ on the test set, as discussed in NetVigil. Although this setup is skewed in favor of these threshold-based methods, it allows us to evaluate the theoretical upper bound of these methods’ performance.

\subsubsection{Implementation}

We implement MalMoE using 900 lines of Python code, based on Pandas~\cite{pandas_web}, PyTorch~\cite{pytorch_web}, and Deep Graph Library (DGL)~\cite{dgl_web}. All experiments are conducted on a single NVIDIA RTX 3090 with 24GiB of GPU memory and a 32-core AMD Ryzen 9 5950X Processor. As for evaluation metrics, we use Accuracy (ACC) and F1-score (F1) as our main metrics.

All flow statistical features are normalized before training, and during testing, we use the same normalization parameters as during training. For hyperparameters, we set $(\alpha_1,\beta_1,\gamma_1)$ to $(0.2, 0.5, 0.5)$ and set $(\alpha_2, \beta_2, \gamma_2)$ to $(0.0, 1.0, 1.0)$.

\subsection{Evaluation of Different Methods}

Table \ref{tab:real} shows the detection accuracy and F1 scores of different methods under different datasets (NF-CSE-CIC-IDS2018-v3, NF-ToN-IoT-v3, NF-UNSW-NB15-v3, NF-BoT-IoT-v3, and our synthetic dataset), \textcolor{blue}{without (w/o)} and \textcolor{red}{with (w/)} graph drift in testing set.

For \textbf{NetBeacon}, it achieves high accuracy on non-drift scenarios (except for the synthetic dataset due to its non-discriminative single-flow features), but due to overfitting, it fails under graph drift.

For \textbf{E-GraphSAGE}, by analyzing the graph structure, it shows superior performance to per-flow analysis (NetBeacon) on non-drift scenarios. But in essence, it is equivalent to using the average traffic feature to represent nodes (just like our Avg-Expert), which renders its failure under flow statistic drift.

For \textbf{HyperVision}, its outlier detection mechanism requires carefully tuned hyperparameters, rendering its performance extremely unstable. For example, on UNSW-NB15, it even achieves higher performance under drifts.

For \textbf{NetVigil}, the anomaly detection mechanism performs well only when the malicious flows are unconnected to benign ones or constitute a small portion of the traffic. Because otherwise, compared with training graphs (which only contain benign flows), malicious flows in the testing graph can alter the context of benign nodes, making them appear anomalous. For example, it achieves the worst metrics on UNSW-NB15, which is predominantly composed of malicious flows.

By combining the robustness of individual experts, MalMoE can achieve the best detection performance under most datasets, either with or without drifts. Also, it achieves the best overall performance, which outperforms the baselines by at least 24\% in ACC and 31\% in F1 under graph drifts.

\subsection{Effectiveness of MoE}
\label{subsec:effectiveness}


Table \ref{tab:synthetic} shows the performance of different variants of MalMoE under different graph drifts, where "Drift 1" refers to flow statistic drift, "Drift 2" refers to graph scale drift, and "Drift 1,2" refers to the coexistence of both drifts.

\textbf{"AVG"} and \textbf{"DEG"} represent the Avg-Expert and the Deg-Expert trained without any data augmentation. They show robustness under their corresponding drifts, but show poor performance under the other, which is consistent with our insight as discussed in Section~\ref{subsec:motivation}.

\textbf{"AVG-AUG"} and \textbf{"DEG-AUG"} represent the augmented versions of the experts. It can be seen that through augmentation, their robustness has been significantly improved. It's because the training and testing datasets are both augmented with flow statistic perturbation and random edge dropping. However, the performance can be further improved by using MoE, as shown by the last row (MalMoE).

We also evaluate a simple variant, \textbf{"AVG-DEG"}, which concatenates the average traffic feature and the node degree feature together as the representation of each node. This manner enforces a single model to handle all the information. However, when a certain drift happens (e.g., Drift 2, graph scale drift), although some of the node information stays stable (e.g., average traffic), the other part varies a lot (e.g., node degree). Even with data augmentation (\textbf{"AVG-DEG-AUG"}), it still fails to outperform MalMoE. It's because, by using MoE, the tasks are assigned to different modules, which eases the burden of individual modules, leading to better robustness.

\textbf{"MalMoE (wo AUG)"} stands for MalMoE but using non-augmented experts (i.e., $(\alpha_1,\beta_1,\gamma_1)=(0,0,0)$). It can achieve much better performance than single experts ("AVG" and "DEG"), except for "Drift 1,2". For "Drift 1,2", the poor performance of experts limits its maximum performance. It can be seen that with augmented experts (\textbf{"MalMoE"}), the performance under "Drift 1,2" improves as well, achieving the best overall performance.

\begin{figure}
    \centering
      \begin{minipage}[t]{0.49\linewidth}
        \centering
        \includegraphics[width=\linewidth]{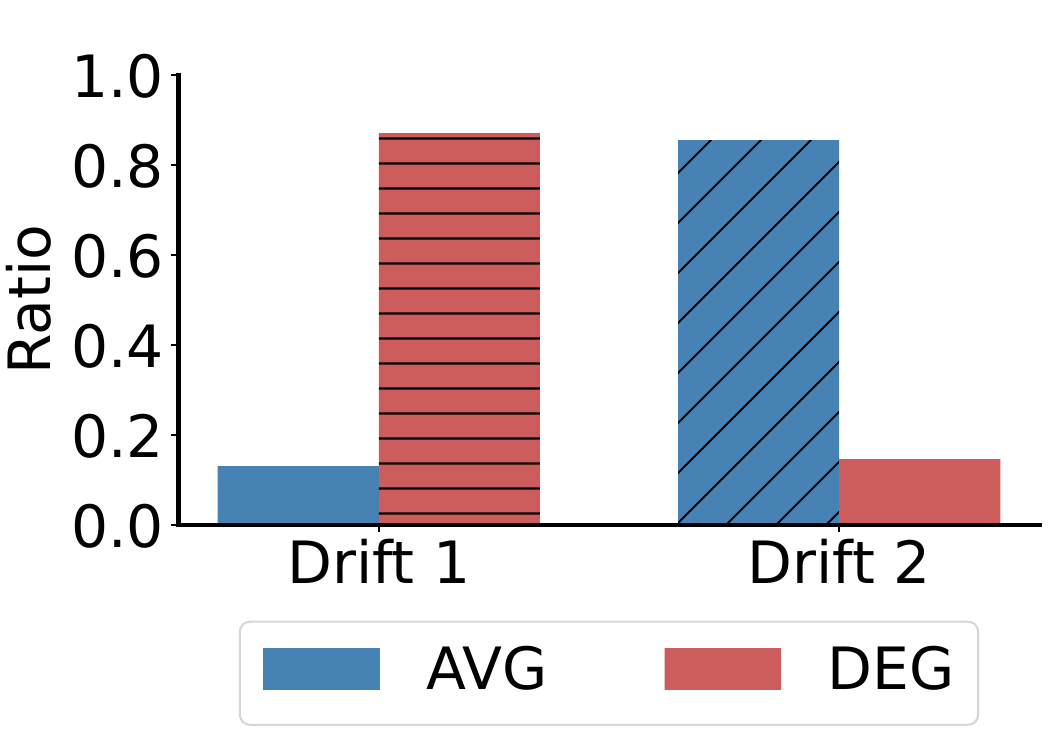}
        \caption{Expert selection distribution on graph drifts.}
        \label{fig:gate_ratio}
      \end{minipage}%
      \hfill
      \begin{minipage}[t]{0.49\linewidth}
        \centering
        \includegraphics[width=\linewidth]{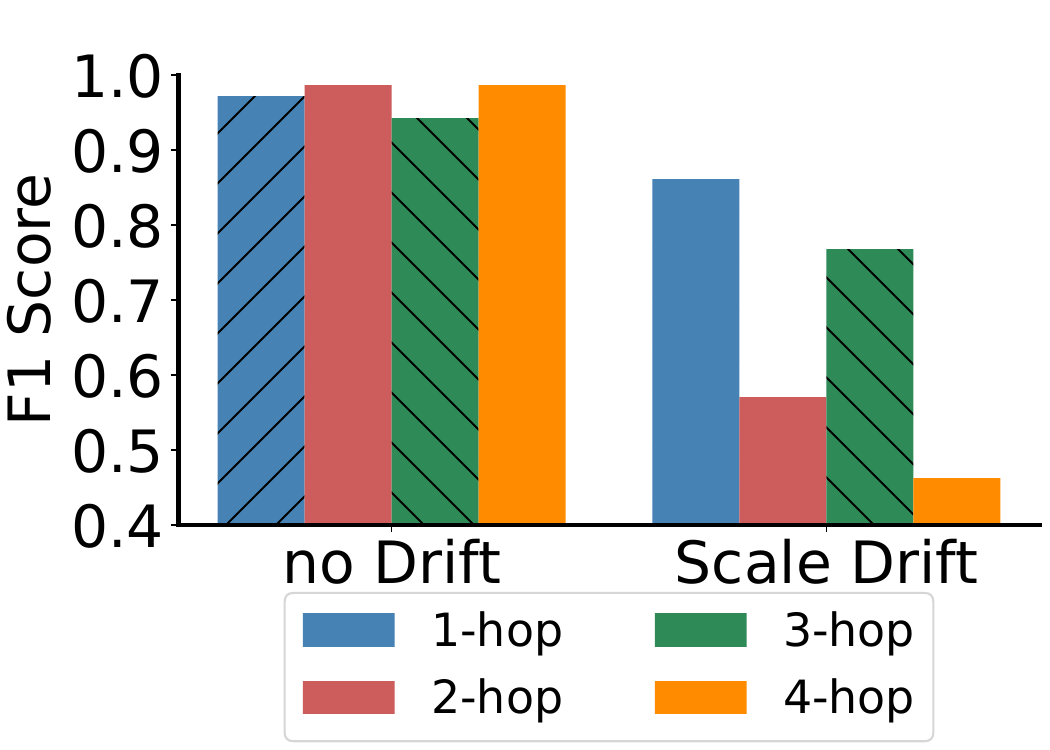}
        \caption{F1 scores of different-hop GCNs with/without drifts.}
        \label{fig:diff_hop}
      \end{minipage}
\end{figure}

To show the explanability of MalMoE, we show the weight assignment of "MalMoE (wo AUG)" in Figure~\ref{fig:gate_ratio}. It illustrates how the gate selects experts on samples where experts show different predictions. For example, the bar of "AVG" on "Drift 1" shows the frequency of selecting Avg-Expert when two experts predict differently. It can be seen that under different graph drifts, the gate model can select the correct expert model, which can also, in turn, reflect the drift distribution of the current test data.

\subsection{Ablation Study}

\begin{table}
    \caption{Results on the synthetic dataset.}
    \label{tab:synthetic}
    \centering
    \setlength{\tabcolsep}{5pt}  
    \renewcommand{\arraystretch}{1.2}  
    \begin{tabular}{c|c|c|c|c|c|c}
        \hline
        Method & Metric & no Drift & Drift 1 & Drift 2 & Drift 1,2 & Overall \\
        \hline
        \multirow{2}{*}{AVG} & ACC & 0.9704 & 0.5882 & 0.8664 & 0.6187 & 0.7609\\
         & F1 & 0.9710 & 0.4822 & 0.8610 & 0.5183 & 0.7081\\
        \hline
        \multirow{2}{*}{DEG} & ACC & 0.9815 & 0.9636 & 0.6122 & 0.5746 & 0.7830\\
         & F1 & 0.9817 & 0.9630 & 0.2811 & 0.2472 & 0.6183\\
        \hline
        \multirow{2}{*}{AVG-AUG} & ACC & 0.9723 & 0.9566 & 0.8740 & 0.8647 & 0.9169\\
         & F1 & 0.9730 & 0.9566 & 0.8536 & 0.8617 & 0.9112\\
        \hline
        \multirow{2}{*}{DEG-AUG} & ACC & 0.9815 & 0.9736 & 0.8869 & 0.8982 & 0.9351\\
         & F1 & 0.9818 & 0.9740 & 0.8623 & 0.8828 & 0.9252\\
        \hline
        \multirow{2}{*}{AVG-DEG} & ACC & 0.9844 & 0.9122 & 0.7523 & 0.7318 & 0.8452\\
         & F1 & 0.9844 & 0.9031 & 0.6618 & 0.6263 & 0.7939\\
        \hline
        AVG-DEG & ACC & 0.9856 & \textbf{0.9835} & 0.8910 & 0.9152 & 0.9438\\
        -AUG & F1 & 0.9858 & \textbf{0.9836} & 0.8492 & 0.9010 & 0.9299\\
        \hline
        MalMoE & ACC & 0.9867 & 0.9380 & \textbf{0.9206} & 0.7188 & 0.8910\\
        (w/o AUG) & F1 & 0.9868 & 0.9370 & \textbf{0.9097} & 0.6018 & 0.8588\\
        \hline
        \multirow{2}{*}{MalMoE} & ACC & \textbf{0.9880} & 0.9774 & 0.9159 & \textbf{0.9194} & \textbf{0.9502}\\
         & F1 & \textbf{0.9881} & 0.9772 & 0.9018 & \textbf{0.9103} & \textbf{0.9444}\\
         \hline
    \end{tabular}
\end{table}

In this part, we evaluate some variants of MalMoE on the synthetic dataset to illustrate the effectiveness of our designs.

\subsubsection{Expert Model Ablation}
\label{subsubsec:expert_ablation}

We use Avg-Expert to show that our 1-hop-GNN-like network design performs the best, as shown in Figure \ref{fig:diff_hop}. As node features already encode 1-hop information, we add multiple GCN layers before the concatenation of node and edge features to capture 2-hop, 3-hop, and 4-hop information. Without any graph drifts, all variants perform well, but under graph scale drift, the robustness of Avg-Expert is severely impaired. It's because the traffic detection usually requires only 1-hop information, whereas multi-hop GNNs aggregate information from too many neighbors, including irrelevant or noisy ones, leading to overfitting. Besides, we found that multi-hop GNNs increase memory consumption to 137\% of the original. Thus, the proposed 1-hop-GNN-inspired design is best suited for our scenario.

\begin{table}[t]
    \caption{Ablation study of gate input and training loss on the synthetic dataset.}
    \label{tab:ablation}
    \centering
    \setlength{\tabcolsep}{4pt}  
    \renewcommand{\arraystretch}{1.2}  
    \begin{tabular}{c|c|c|c|c|c|c}
        \hline
        Method & Metric & no Drift & Drift 1 & Drift 2 & Drift 1,2 & Overall \\
        \hline
        \multirow{2}{*}{w/o G.I.} & $\mathrm{ACC}_{cls}$ & 0.9878 & 0.9744 & 0.9096 & 0.9171 & 0.9472\\
        & $\mathrm{ACC}_{gate}$ & 0.9418 & 0.7805 & 0.7702 & 0.7754 & 0.8170\\
         \hline
        \multirow{2}{*}{w/o H.S.} & $\mathrm{ACC}_{cls}$ & 0.9875 & \textbf{0.9863} & 0.8745 & 0.8981 & 0.9366\\
        & $\mathrm{ACC}_{gate}$ & 0.3169 & 0.3477 & 0.1366 & 0.1705 & 0.2429\\
         \hline
         \multirow{2}{*}{w/o AUG} & $\mathrm{ACC}_{cls}$ & 0.9877 & 0.9727 & \textbf{0.9207} & 0.9085 & 0.9474\\
        & $\mathrm{ACC}_{gate}$ & 0.9388 & 0.7562 & 0.7913 & 0.6742 & 0.7901\\
         \hline
         \multirow{2}{*}{One-Stage} & $\mathrm{ACC}_{cls}$ & 0.9879 & 0.9800 & 0.9031 & 0.9183 & 0.9474\\
        & $\mathrm{ACC}_{gate}$ & 0.9434 & \textbf{0.8548} & 0.7619 & 0.7734 & 0.8334\\
         \hline
        \multirow{2}{*}{MalMoE} & $\mathrm{ACC}_{cls}$ & \textbf{0.9880} & 0.9774 & 0.9159 & \textbf{0.9194} & \textbf{0.9502}\\
        & $\mathrm{ACC}_{gate}$ & \textbf{0.9529} & 0.8301 & \textbf{0.8023} & \textbf{0.7821} & \textbf{0.8419}\\
         \hline
    \end{tabular}
\end{table}

\subsubsection{Gate Model Ablation}
\label{subsubsec:gate_ablation}

We show the influence of the gate input and the gate output, as shown in Table~\ref{tab:ablation}. We evaluate two metrics: $\mathrm{ACC}_{cls}$ is the accuracy of the benign/malicious classification, which represents the overall performance, and $\mathrm{ACC}_{gate}$ is the accuracy of gating, which represents the explainability of the gating weights.

Without using graph representation as input (\textbf{"w/o G.I."}), the gate makes decisions only based on per-sample information. In this case, sometimes it's nearly impossible for the gate to judge whether a sample is near-OOD, which reduces the overall $\mathrm{ACC}_{gate}$ from 0.8419 to 0.8170.

When using the "weighted summation" mechanism like traditional MoEs instead of hard selection (\textbf{"w/o H.S."}), the latent vectors of each expert model are weighted by the gating outputs and summed up, as the input to a classification head. On the one hand, the additional classification head increases the training difficulty of the model, leading to a decrease in $\mathrm{ACC}_{cls}$ (from 0.9502 to 0.9366). On the other hand, different expert models don't share the same latent space, which renders the gating weights meaningless. This leads to a significant decrease of $\mathrm{ACC}_{gate}$ (from 0.8419 to 0.2429).

\subsubsection{Training Strategy Ablation}
\label{subsubsec:training_ablation}

We evaluate the influence of the data augmentation module and the two-stage training strategy, as shown in Table \ref{tab:ablation}.

By disabling the data augmentation during gate training(\textbf{"w/o AUG"}, $(\alpha_2,\beta_2,\gamma_2)=(0,0,0)$), the gate learns how to route solely from non-drift data, where both experts perform equally well, which reduces $\mathrm{ACC}_{gate}$ from 0.8419 to 0.7901.

Training the experts and gate together (\textbf{"One-Stage"}) only causes a minor impact. This might be because the design of the gate model is simple, allowing it to quickly adapt to changes in the experts' performance. But it still slightly harms the performance of the gate.

\subsection{Real-world Traces}

\begin{figure}
    \centering
    \includegraphics[width=\linewidth]{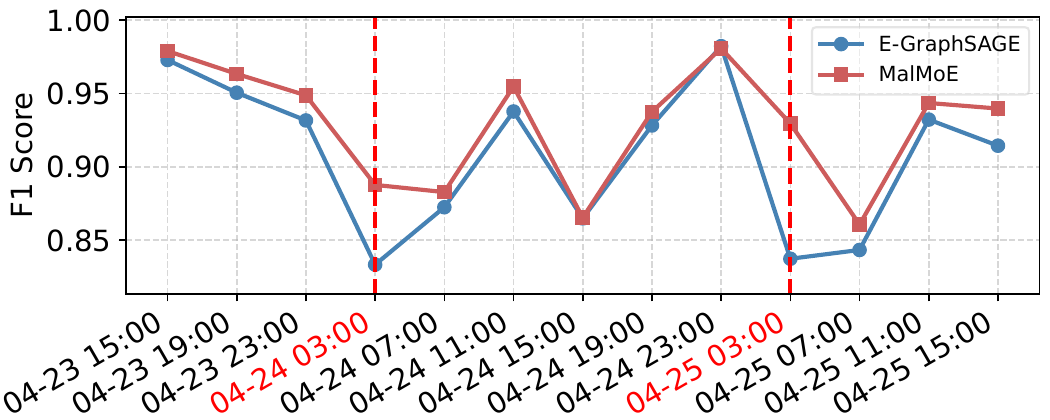}
    \caption{Performance of E-GraphSAGE and MalMoE as time varies on the real-world trace.}
    \label{fig:cernet}
\end{figure}

To validate the practical utility of MalMoE, we test it with real-world NetStream~\cite{netstream_web} traces from a well-known backbone network operator, in comparison with E-GraphSAGE (which can be seen as Avg-Expert). We set the time window for graph construction to 30 seconds, which consists of about 200,000 flows. For both models, we use 6 graphs from [2025-04-23 15:00:00, 2025-04-23 15:02:30] for training, and use the subsequent 48 hours of data for testing, as shown in Figure~\ref{fig:cernet}.

The performance of E-GraphSAGE shows certain periodicity. It's because the traffic within a single day forms a cycle as shown in Section \ref{subsec:challenges}. Therefore, the farther the testing clock time is from the training clock time, the worse the performance is expected to be. That's why E-GraphSAGE achieves the lowest F1 scores at 04-24 03:00 and 04-25 03:00, which is the farthest clock time from the training clock time (15:00). But MalMoE can improve the F1 scores at these two time points from 0.8334 to 0.8877, and from 0.8374 to 0.9295, respectively. There's an exception that the F1 drops at 04-24 15:00, which means that there are also differences in traffic distribution at the same clock time across different days.

\subsection{Efficiency Evaluation}

We apply the MAWI dataset to estimate the efficiency of MalMoE. In our initial implementation, we found that the time bottleneck lies in graph construction, and the memory bottleneck lies in graph processing. For the graph construction, the slowest operation is mapping IP strings to node indices. We accelerate it by using Hash Mapping~\cite{pandas_documentation} to avoid sorting. For the graph processing, MLP inference consumes the most GPU memory. We reduce the consumption by using batched inference instead of inputting the entire graph at once. Derived from Table \ref{tab:efficiency}, after optimization, we can process 858,646 flows/s, and only consumes 1.718 GB/Mflow. This enables real-time detection in high-volume scenarios like ISP networks.

\section{Related Work}

\subsection{Graph-Based Malicious Flow Detection}

Some works utilize graph information to model flow interactions~\cite{hsieh2024netvigil,lo2022graphsage,fu2023detecting,nguyen2023ts,duan2024practical,altaf2023ne,duan2022application} to detect malicious traffic that exhibits similar behavior to benign traffic. E-GraphSAGE~\cite{lo2022graphsage} is the first to apply GNN for flow-level malicious traffic detection, but it doesn't consider temporal graph drifts at all. DLGNN~\cite{duan2022application} constructs line graphs (flows as nodes), and uses GNN to analyze spatial information and Gated Recurrent Unit (GRU) to analyze temporal information. But it ignores that the time-series information varies with time as well. Compared to these works, MalMoE explicitly addresses the issue of temporal graph drift by leveraging robust node features.

\subsection{Drift-Aware Flow Analysis}
\label{subsec:drift_aware_flow_analysis}

Existing detection methods considering traffic drift use either unsupervised learning, retraining, or data augmentation. Whisper~\cite{fu2021realtime} extracts the frequency features for each flow and applies clustering-based anomaly detection. But when a benign flow to be detected is also far from the training cluster centers, it will be mistakenly classified as malicious. CD-Net~\cite{chen2025cd} applies a CNN for representation learning and a DNN for classification, and during detection, it finetunes the DNN to handle drifts. However, in reality, labeling can only be done by network security experts, and the labels are also somewhat imprecise. Rosetta~\cite{xie2023rosetta} applies different TCP-aware traffic augmentation mechanisms to enable dynamic detection. But its traffic extractor is a single flow-level model, which cannot handle the drifts beyond the TCP-related ones. For graph drifts, HyperVision~\cite{fu2023detecting} applies an unsupervised way, and NetVigil~\cite{hsieh2024netvigil} applies all methods above, so they both suffer from the shortcomings. Compared to these works, MalMoE combines multiple models that utilize different inherently robust node feature types, which avoids costly retraining and over-reliance on a single model/representation.

\subsection{Mixture of Experts}

\begin{table}
\caption{Estimation of time and memory on MAWI.}
\label{tab:efficiency}
\centering
\begin{threeparttable}
    \setlength{\tabcolsep}{6pt}  
    \renewcommand{\arraystretch}{1.2}  
    \begin{tabular}{c|c|c}
        \hline
        Stage & Time(s) & Memory(GB)\\
        \hline
        Construction & 6.08 & 0.283\\
        \hline
        Processing & 1.22 & 10.77\\
        \hline
    \end{tabular}
\end{threeparttable}
\end{table}

There have been tons of MoE-based works, which mainly focus on improving the model capacity or combining the experts' strengths in their domains. Some of them are related to MalMoE. SNAKE~\cite{qin2025snake} applies MoE to address a set of network traffic classification tasks, but it doesn't consider the graph drifts. Hierarchical MoE~\cite{li2025hierarchical} uses MoE to integrate multiple granularities of a graph to make HLS prediction, but their experts take care of different granularities instead of different drifts. GraphMETRO~\cite{wu2024graphmetro} also makes different experts tackle different graph drifts, but their experts all share the same node features, edge features, and model structure, which is essentially incapable of handling certain types of drift. Besides, it still applies the traditional MoE structure and doesn't consider the generalizability and explainability of the gate. Compared to these works, MalMoE is the first to use MoE for combining the robustness of different node features.

\section{Future Work and Conclusion}

MalMoE can be further improved in two directions. First, its extensible design allows incorporating additional node-feature types—e.g., maximum-traffic representations or flow-category representations~\cite{tan2025graph}—to handle a wider range of graph drifts.
Second, the expert of MalMoE can use more complex graph analysis methods, such as Graph Attention Networks (GAT) and Graph Isomorphism Network (GIN).


In this paper, we propose MalMoE, a retraining-free, extensible, real-time system for encrypted traffic detection under temporal graph drifts. MalMoE leverages an MoE framework to fuse the complementary drift robustness of different node-feature types: we build simple yet effective experts for edge-level classification, redesign the gate’s inputs/outputs for drift-aware routing, and adopt a two-stage training strategy with augmentation for stable learning. Experiments on open-source, synthetic, and real-world datasets demonstrate strong effectiveness and efficiency, and we hope MalMoE offers useful insights toward addressing drift in traffic analysis.

\section*{Acknowledgment}

This work was sponsored by the NSFC grant(62431017). We gratefully acknowledge the support of Key Laboratory of Intelligent Press Media Technology.

\bibliographystyle{IEEEtran}
\bibliography{IEEEabrv,mybibfile}

@inproceedings{hsieh2024netvigil,
  title={$\{$NetVigil$\}$: Robust and $\{$Low-Cost$\}$ Anomaly Detection for $\{$East-West$\}$ Data Center Security},
  author={Hsieh, Kevin and Wong, Mike and Segarra, Santiago and Mani, Sathiya Kumaran and Eberl, Trevor and Panasyuk, Anatoliy and Netravali, Ravi and Chandra, Ranveer and Kandula, Srikanth},
  booktitle={21st USENIX Symposium on Networked Systems Design and Implementation (NSDI 24)},
  pages={1771--1789},
  year={2024}
}

@inproceedings{lo2022graphsage,
  title={E-graphsage: A graph neural network based intrusion detection system for iot},
  author={Lo, Wai Weng and Layeghy, Siamak and Sarhan, Mohanad and Gallagher, Marcus and Portmann, Marius},
  booktitle={NOMS 2022-2022 IEEE/IFIP Network Operations and Management Symposium},
  pages={1--9},
  year={2022},
  organization={IEEE}
}

@inproceedings{sarhan2021netflow,
  title={Netflow datasets for machine learning-based network intrusion detection systems},
  author={Sarhan, Mohanad and Layeghy, Siamak and Moustafa, Nour and Portmann, Marius},
  booktitle={Big data technologies and applications: 10th EAI international conference, BDTA 2020, and 13th EAI international conference on wireless internet, WiCON 2020, virtual event, December 11, 2020, proceedings 10},
  pages={117--135},
  year={2021},
  organization={Springer}
}

@article{fu2023detecting,
  title={Detecting unknown encrypted malicious traffic in real time via flow interaction graph analysis},
  author={Fu, Chuanpu and Li, Qi and Xu, Ke},
  journal={arXiv preprint arXiv:2301.13686},
  year={2023}
}

@article{nguyen2023ts,
  title={TS-IDS: Traffic-aware self-supervised learning for IoT Network Intrusion Detection},
  author={Nguyen, Hoang and Kashef, Rasha},
  journal={Knowledge-Based Systems},
  volume={279},
  pages={110966},
  year={2023},
  publisher={Elsevier}
}

@article{duan2024practical,
  title={Practical cyber attack detection with continuous temporal graph in dynamic network system},
  author={Duan, Guanghan and Lv, Hongwu and Wang, Huiqiang and Feng, Guangsheng and Li, Xiaoli},
  journal={IEEE Transactions on Information Forensics and Security},
  year={2024},
  publisher={IEEE}
}

@article{luo2024identifying,
  title={Identifying malicious traffic under concept drift based on intraclass consistency enhanced variational autoencoder},
  author={Luo, Xiang and Liu, Chang and Gou, Gaopeng and Xiong, Gang and Li, Zhen and Fang, Binxing},
  journal={Science China Information Sciences},
  volume={67},
  number={8},
  pages={182302},
  year={2024},
  publisher={Springer}
}

@article{chen2025cd,
  title={CD-Net: Robust mobile traffic classification against apps updating},
  author={Chen, Yanan and Hou, Botao and Wu, Bin and Hu, Hao},
  journal={Computers \& Security},
  volume={150},
  pages={104214},
  year={2025},
  publisher={Elsevier}
}

@inproceedings{xavier2024fast,
  title={Fast Learning Enabled by In-Network Drift Detection},
  author={Xavier, Bruno Missi and Martinello, Magnos and Trois, Celio and Alenca, Brenno M and Rios, Ricardo A},
  booktitle={Proceedings of the 8th Asia-Pacific Workshop on Networking},
  pages={129--134},
  year={2024}
}

@inproceedings{deng2024robust,
  title={Robust and reliable early-stage Website fingerprinting attacks via spatial-temporal distribution analysis},
  author={Deng, Xinhao and Li, Qi and Xu, Ke},
  booktitle={Proceedings of the 2024 on ACM SIGSAC Conference on Computer and Communications Security},
  pages={1997--2011},
  year={2024}
}

@inproceedings{el2017survey,
  title={A survey on deep packet inspection},
  author={El-Maghraby, Reham Taher and Abd Elazim, Nada Mostafa and Bahaa-Eldin, Ayman M},
  booktitle={2017 12th International Conference on Computer Engineering and Systems (ICCES)},
  pages={188--197},
  year={2017},
  organization={IEEE}
}

@inproceedings{hu2007malware,
  title={A malware signature extraction and detection method applied to mobile networks},
  author={Hu, Guoning and Venugopal, Deepak},
  booktitle={2007 IEEE International Performance, Computing, and Communications Conference},
  pages={19--26},
  year={2007},
  organization={IEEE}
}

@inproceedings{ssl2015data,
  title={Data Mining Approach for Detection of DDoS},
  author={SSL, Attacks Utilizing and Protocol, TLS},
  booktitle={Internet of Things, Smart Spaces, and Next Generation Networks and Systems: 15th International Conference, NEW2AN 2015, and 8th Conference, ruSMART 2015, St. Petersburg, Russia, August 26-28, 2015, Proceedings},
  volume={9247},
  pages={274},
  year={2015},
  organization={Springer}
}

@inproceedings{qin2015ddos,
  title={DDoS attack detection using flow entropy and clustering technique},
  author={Qin, Xi and Xu, Tongge and Wang, Chao},
  booktitle={2015 11th International Conference on Computational Intelligence and Security (CIS)},
  pages={412--415},
  year={2015},
  organization={IEEE}
}

@inproceedings{petliak2023signature,
  title={Signature-based approach to detecting malicious outgoing traffic.},
  author={Petliak, Nataliia and Klots, Yurii and Titova, Vira and Cheshun, Viktor and Boyarchuk, Artem},
  booktitle={IntelITSIS},
  pages={486--506},
  year={2023}
}

@misc{SSL_Certificate_Authorities,
  author = {Web Technology Surveys},
  title = {Usage statistics and market shares of SSL certificate authorities for websites},
  year = {2025},
  url = {https://w3techs.com/technologies/overview/ssl_certificate},
  note = {2025-05-30}
}

@misc{ThreatLabz_2024_Encrypted_Attacks_Report,
  author = {Zscaler},
  title = {ThreatLabz 2024 Encrypted Attacks Report},
  year = {2024},
  url = {https://www.zscaler.com/campaign/threatlabz-encrypted-attacks-report},
  note = {2025-05-31}
}

@misc{luay2025NetFlowDatasetsV3,
title = {Temporal Analysis of NetFlow Datasets for Network Intrusion Detection Systems},
author = {Majed Luay and Siamak Layeghy and Seyedehfaezeh Hosseininoorbin and Mohanad Sarhan and Nour Moustafa and Marius Portmann},
year = {2025},
eprint = {2503.04404},
archivePrefix= {arXiv},
primaryClass = {cs.LG},
url = {https://arxiv.org/abs/2503.04404}
}

@article{qin2025snake,
  title={SNAKE: A Sustainable and Multi-functional Traffic Analysis System utilizing Specialized Large-Scale Models with a Mixture of Experts Architecture},
  author={Qin, Tian and Cheng, Guang and Zhou, Yuyang and Chen, Zihan and Luan, Xing},
  journal={arXiv preprint arXiv:2503.13808},
  year={2025}
}

@inproceedings{li2025hierarchical,
  title={Hierarchical mixture of experts: Generalizable learning for high-level synthesis},
  author={Li, Weikai and Wang, Ding and Ding, Zijian and Sohrabizadeh, Atefeh and Qin, Zongyue and Cong, Jason and Sun, Yizhou},
  booktitle={Proceedings of the AAAI Conference on Artificial Intelligence},
  volume={39},
  number={17},
  pages={18476--18484},
  year={2025}
}

@article{wu2024graphmetro,
  title={GraphMETRO: Mitigating Complex Graph Distribution Shifts via Mixture of Aligned Experts},
  author={Wu, Shirley and Cao, Kaidi and Ribeiro, Bruno and Zou, James Y and Leskovec, Jure},
  journal={Advances in Neural Information Processing Systems},
  volume={37},
  pages={9358--9387},
  year={2024}
}

@misc{pandas_documentation,
  author = {pandas},
  title = {pandas 2.2.3 documentation},
  year = {2024},
  url = {https://pandas.pydata.org/docs/reference/api/pandas.factorize.html},
  note = {2025-05-31}
}

@misc{netflow_documentation,
  author = {Cisco},
  title = {Cisco IOS NetFlow},
  year = {2025},
  url = {https://www.cisco.com/c/en/us/products/ios-nx-os-software/ios-netflow/index.html},
  note = {2025-05-31}
}

@misc{mawi_documentation,
  author = {MAWI Working Group},
  title = {MAWI Working Group Traffic Archive},
  year = {2025},
  url = {https://mawi.wide.ad.jp/mawi/},
  note = {2025-05-31}
}

@inproceedings{zhou2023efficient,
  title={An efficient design of intelligent network data plane},
  author={Zhou, Guangmeng and Liu, Zhuotao and Fu, Chuanpu and Li, Qi and Xu, Ke},
  booktitle={32nd USENIX Security Symposium (USENIX Security 23)},
  pages={6203--6220},
  year={2023}
}

@misc{dgl_web,
  author = {Deep Graph Library Committors},
  title = {Deep Graph Library},
  year = {2025},
  url = {https://www.dgl.ai/},
  note = {2025-07-04}
}

@misc{pytorch_web,
  author = {Pytorch Team},
  title = {Pytorch},
  year = {2025},
  url = {https://pytorch.org/},
  note = {2025-07-04}
}

@misc{pandas_web,
  author = {pandas volunteer contributors},
  title = {pandas},
  year = {2025},
  url = {https://pandas.pydata.org/},
  note = {2025-07-04}
}

@inproceedings{wichtlhuber2022ixp,
  title={Ixp scrubber: learning from blackholing traffic for ml-driven ddos detection at scale},
  author={Wichtlhuber, Matthias and Strehle, Eric and Kopp, Daniel and Prepens, Lars and Stegmueller, Stefan and Rubina, Alina and Dietzel, Christoph and Hohlfeld, Oliver},
  booktitle={Proceedings of the ACM SIGCOMM 2022 Conference},
  pages={707--722},
  year={2022}
}

@misc{netstream_web,
  author = {Liu Jieyuan},
  title = {What Is NetStream?},
  year = {2025},
  url = {https://info.support.huawei.com/info-finder/encyclopedia/en/NetStream.html},
  note = {2025-07-22}
}

@article{jiang2024med,
  title={Med-moe: Mixture of domain-specific experts for lightweight medical vision-language models},
  author={Jiang, Songtao and Zheng, Tuo and Zhang, Yan and Jin, Yeying and Yuan, Li and Liu, Zuozhu},
  journal={arXiv preprint arXiv:2404.10237},
  year={2024}
}

@article{patsakis2020encrypted,
  title={Encrypted and covert DNS queries for botnets: Challenges and countermeasures},
  author={Patsakis, Constantinos and Casino, Fran and Katos, Vasilios},
  journal={Computers \& Security},
  volume={88},
  pages={101614},
  year={2020},
  publisher={Elsevier}
}

@inproceedings{bumanglag2020impact,
  title={On the impact of DNS over HTTPS paradigm on cyber systems},
  author={Bumanglag, Kimo and Kettani, Houssain},
  booktitle={2020 3rd International Conference on Information and Computer Technologies (ICICT)},
  pages={494--499},
  year={2020},
  organization={IEEE}
}

@article{altaf2023ne,
  title={NE-GConv: A lightweight node edge graph convolutional network for intrusion detection},
  author={Altaf, Tanzeela and Wang, Xu and Ni, Wei and Liu, Ren Ping and Braun, Robin},
  journal={Computers \& Security},
  volume={130},
  pages={103285},
  year={2023},
  publisher={Elsevier}
}

@article{duan2022application,
  title={Application of a dynamic line graph neural network for intrusion detection with semisupervised learning},
  author={Duan, Guanghan and Lv, Hongwu and Wang, Huiqiang and Feng, Guangsheng},
  journal={IEEE Transactions on Information Forensics and Security},
  volume={18},
  pages={699--714},
  year={2022},
  publisher={IEEE}
}

@article{jacobs1991adaptive,
  title={Adaptive mixtures of local experts},
  author={Jacobs, Robert A and Jordan, Michael I and Nowlan, Steven J and Hinton, Geoffrey E},
  journal={Neural computation},
  volume={3},
  number={1},
  pages={79--87},
  year={1991},
  publisher={MIT Press}
}

@inproceedings{tan2025graph,
  title={Graph-Based Encrypted Malicious Traffic Detection Under Flow Distribution Drift With Flow Sampling},
  author={Tan, Yunpeng and Li, Qingyang and Yang, Mingxin and Zhang, Xinggong},
  booktitle={Proceedings of the 9th Asia-Pacific Workshop on Networking},
  pages={261--262},
  year={2025}
}

@inproceedings{fu2021realtime,
  title={Realtime robust malicious traffic detection via frequency domain analysis},
  author={Fu, Chuanpu and Li, Qi and Shen, Meng and Xu, Ke},
  booktitle={Proceedings of the 2021 ACM SIGSAC Conference on Computer and Communications Security},
  pages={3431--3446},
  year={2021}
}

@inproceedings{xie2023rosetta,
  title={Rosetta: Enabling robust tls encrypted traffic classification in diverse network environments with tcp-aware traffic augmentation},
  author={Xie, Renjie and Wang, Yixiao and Cao, Jiahao and Dong, Enhuan and Xu, Mingwei and Sun, Kun and Li, Qi and Shen, Licheng and Zhang, Menghao},
  booktitle={Proceedings of the ACM turing award celebration conference-China 2023},
  pages={131--132},
  year={2023}
}

\end{document}